\documentclass[%
	,preprint
	,nofootinbib
	,amsmath
	,amssymb
	,aps
	,a4paper
]{revtex4-1}
\PassOptionsToPackage{log-declarations=false}{xparse}
\usepackage{graphicx}
\makeatletter
\let\MYcaption\@makecaption
\usepackage[
	,font=footnotesize
]{subcaption}
\let\@makecaption\MYcaption
\makeatother
\usepackage{xcolor}
\PassOptionsToPackage{log-declarations=false}{xparse}
\usepackage{physics}

\usepackage{mleftright}\mleftright
\usepackage[T1]{fontenc}
\usepackage{lmodern}
\usepackage{libertineRoman}
\usepackage[
	,slantedGreek
	,libertine
	,libaltvw
	,liby
]{newtxmath}
\usepackage{bm}
\usepackage[
	,unicode
	,colorlinks
	,allcolors=blue!90!cyan!80!black!90!white
	,setpagesize=false
	,bookmarksnumbered=true
	,bookmarksopen=true
	,pdfborder={0 0 0}
]{hyperref}
\newcommand{\ext}{{\mathrm{ext}}}

\newcommand{\F}{\mathrm{F}}

\newcommand\B{\mathrm{B}}
\newcommand{\BB}{\mathrm{BB}}
\newcommand{\BF}{\mathrm{BF}}

\newcommand{\br}{{\bm r}}
\newcommand{\bp}{{\bm p}}
\newcommand{\bq}{{\bm q}}
\newcommand{\kF}{k_{\mathrm F}}
\newcommand{\kB}{k_{\mathrm B}}
\newcommand{\ui}{{\mathrm i}}
\newcommand{\ue}{{\mathrm e}}
\newcommand{\cI}{{\mathcal I}}
\newcommand{\wf}{{\widetilde f}}
\newcommand{\veps}{\varepsilon}
\newcommand{\vepsF}{\veps_\mathrm{F}}
\newcommand{\BEC}{\mathrm{BEC}}
\begin{document}
\title{\Large Collective Excitations in Bose--Fermi Mixtures}
\author{%
Yoji~Asano, 
Masato~Narushima, 
Shohei~Watabe, and 
Tetsuro~Nikuni
}
\affiliation{%
Department of Physics,
Tokyo University of Science,
1-3 Kagurazaka,
Shinjuku-ku,
Tokyo,
162-8601,
Japan\\
}
\date{\today}

\begin{abstract}
We investigate collective excitations of density fluctuations and a dynamic density structure factor in a mixture of Bose and Fermi gases in a normal phase. With decreasing temperature, we find that the frequency of the collective excitation deviates from that of the hydrodynamic sound mode. Even at a temperature much lower than the Fermi temperature, the collective mode frequency does not reach the collisionless limit analogous to zero sound in a Fermi gas, because of collisions between bosons and fermions.
\end{abstract}
\maketitle

\section{Introduction}
Ultracold atomic quantum gases provide intriguing fields of studying versatile phenomena of Bose--Fermi mixtures~\cite{onofrio_2016_PU_59_physics}, such as heteronuclear molecules~\cite{roy_2016_PRA_94_photoassociative}, Bose and Fermi polarons~\cite{scazza_2017_PRL_118_repulsive}, degenerate Fermi gas trapped by a Bose--Einstein condensate~\cite{desalvo_2017_PRL_119_observation}, and superfluid mixtures~\cite{yao_2016_PRL_117_observation,roy_2017_PRL_118_two}. 
In ultracold atomic gases, a variety of the collective excitations has been studied on, for example,
monopole mode of a degenerate Bose gas,
\cite{straatsma_2016_PRA_94_collapse},
first and second sound in a Bose--Einstein condensate,
\cite{arahata_2013_PRA_87_propagation},
zero and first sound in a normal Fermi gas
\cite{watabe_2010_JLTP_158_zero,narushima_2018_JPBAMOP_51_density},
second sound in a superfluid Fermi gas
\cite{sidorenkov_2013_N_498_second,hu_2014_NJP_16_first},
sound modes of a Bose--Fermi mixture superfluid
\cite{yao_2016_PRL_117_observation}.
Experimental cooling and thermometry of Bose--Fermi mixtures are summarised by Ref.~\cite{onofrio_2016_PU_59_physics}.
However, collective excitations in a normal Bose--Fermi mixtures have not been studied intensively and extensively. 
In Bose--Fermi mixtures, since effects of Bose-enhancement and Pauli-blocking simultaneously emerge, there arises the question which property dominates the collective excitation, and how both quantum statistics affect the collective excitation. 
One can expect interesting features of collective excitation in those Bose--Fermi mixtures, caused by completely different features of quantum statistics.

In this paper, we study the density collective excitation in the normal Bose--Fermi mixture gas, and by tuning temperature, we explore new feature of the collective excitation in normal states, which is different from the hydrodynamic first sound and the collisionless zero sound.

\section{Moment Method}
\label{sec:moment_method}
We consider a mixture gas in the normal state composed of the single-component bosonic atoms and single-component fermionic atoms. 
Bosonic atoms with an atomic mass $m_{\B}$ interact with each other with the contact interaction $g_\BB$. 
Fermionic atoms with an atomic mass $m_\F$ do not interact with each other in the case of $s$-wave contact interaction because of the Pauli-blocking, and a ferminic atom only interact with a bosonic atom with the contact interaction $g_\BF$. 
The Boltzmann equations for the bosonic and fermionic distribution functions, $f_\B=f_\B(\bp,\br,t)$ and $f_\F=f_\F(\bp,\br,t)$, are given by 
\begin{subequations}
\label{eq:boltzmann_equation:1}
\begin{align}
&
\pdv{f_\B}{t}+\pdv{\veps_\B}{\bp}\cdot\pdv{f_\B}{\br}
-\pdv{U_\B}{\br}
\cdot\pdv{f_\B}{\bp} 
=
\cI_\B
\;,\\&
\pdv{f_\F}{t}+\pdv{\veps_\F}{\bp}\cdot\pdv{f_\F}{\br}
-\pdv{U_\F}{\br}
\cdot \pdv{f_\F}{\bp} 
=\cI_\F
\;,
\end{align}
\end{subequations}
where the single-particle energies are given by $\veps_{\B,\F}=p^2/2m_{\B,\F}+U_{\B,\F}$. 
The potential terms $U_{\B, \F}$ include the mean-field interaction, given by $U_\B \equiv  g_\BF n_\F+2g_\BB n_\B+U_\B^\ext$ and $U_\F = g_\BF n_\B+U_\F^\ext$, where $U_{\B,\F}^\ext=U_{\B,\F}^\ext(\br,t)$ is the time-dependent external potentials, and $n_{\B,\F}=n_{\B,\F}(\br,t)$ are the local number density of the particles. 
The right-hand sides of the Boltzmann equations $\cI_\B$ and $\cI_\F$ are collision integrals
\cite{abrikosov_1959_RPP_22_theory,hamel_1965_TPoF_8_kinetic,capuzzi_2004_PRA_70_effects,kadanoff_1962_WBI_quantum}.
The static equilibrium solution is given by the Bose--Einstein and Fermi--Dirac distribution functions
$
f^0_{\B,\F}=\bqty*{\smash{\ue^{\pqty*{\veps^0_{\B,\F}-\mu_{\B,\F}}/\kB T}}\mp1}^{-1}
$.
The superscript ``$0$'' indicates that the variables are in static equilibrium.

We linearize the Boltzmann equations around static equilibrium, writing the fluctuations of the distribution functions as
$
f_{\B,\F}-f^0_{\B,\F}\equiv\pqty{\pdv*{\smash{f^0_{\B,\F}}}{\smash{\veps^0_{\B,\F}}}}\nu_{\B,\F}
$.
For the collision integrals, we adopt the relaxation time approximation in the simplest form
\label{eq:average_relaxation_time:1}
\begin{alignat}{2}
&
\cI_\B
=
-\smash{\frac{f_\B-\wf_\B}{\tau_\B}}
\;,
&\quad&
\cI_\F
=-\smash{\frac{f_\F-\wf_\F}{\tau_\F}}
\;. 
\end{alignat}
Here, $\wf_{\B,\F}=\wf_{\B,\F}(\bp,\br,t)$ denotes a local equilibrium distribution function.
The relaxation time $\tau_\B$ contains two contributions of both intraspecies and interspecies collisions, given by $\tau_\B^{-1}=\tau_\BB^{-1}+\tau_\BF^{-1}$, where $\tau_\BB$ and $\tau_\BF$ are the mean-collision times for boson--boson, and boson--fermion collisions, respectively. 
On the other hand, the relaxation time $\tau_\F$ only contains a contribution of the interspecies collision, i.e., $\tau_\F=\tau_\BF$, because of the Pauli-blocking.
For explicit expressions for the collisional relaxation times, see Refs \cite{pethick_2002_Cup_bose,goldwin_2005_PRA_71_cross}.

We consider an external potential that excites modes of frequency $\omega$ and wavevector $\bq$, namely $U^\ext_{\B,\F}(\br,t)=U^\ext_{\B,\F}(\bq,\omega)\ue^{\ui(\bq\cdot\br-\omega t)}$.
We then look for the plane-wave solutions, such as
\begin{alignat}{2}
&
A(\br,t)=A(\bq,\omega)\ue^{\ui(\bq\cdot\br-\omega t)}
\;,
&\quad&
B(\bp,\br,t)=B(\bp,\bq,\omega)\ue^{\ui(\bq\cdot\br-\omega t)}
\;,
\end{alignat}
and expand the distribution function in terms of the spherical harmonics function
(we omit the dependences on $\bq$ and $\omega$ for conciseness):
\begin{align}
\nu_{\B,\F}(\bp)=\sum_{l=0}^\infty\sum_{m=-l}^l\nu_{\B,\F;\,l}^m(p)P_l^m(\cos\theta)\ue^{\ui m\phi}
\;.
\end{align}
In the case of $s$-wave interaction, modes with $m\neq0$ are uncoupled to density fluctuations.
We thus choose the mode $m=0$. 
We make use of the orthogonality of the spherical harmonics and derive a coupled set of equations for the moments
\begin{align}
\ev*{p^n\nu_{\B,\F;\,l}}=\int\mspace{-8mu}\frac{\dd[3]{p}}{(2\uppi\hbar)^{3}}\pdv{f^0_{\B,\F}}{\veps_{\B,\F}}p^n\nu_{\B,\F;\,l}(p)
\;,
\end{align}
where $\nu_{\B,\F;\,l}(p)\equiv\nu_{\B,\F;\,l}^{m=0}(p)$.
In particular, the zeroth moment $\ev*{\nu_{\B,\F;\,0}}$ is equal to the density deviation $\delta n_{\B,\F}(\bq,\omega)$ from the static equilibrium value.
Solving the coupled moment equations in the case $U_\B^\ext=U_\F^\ext=U^\ext$, we obtain the density fluctuation $\delta n=\delta n_\B+\delta n_\F$ in the form $\delta n(\bq,\omega)=\chi_\mathrm{d}(\bq,\omega)U^\ext(\bq,\omega)$, which defines the density response function and hence, the dynamic density structure function is given by
\begin{align}
S_\mathrm{d}(\bq,\omega)
=-\frac1{\uppi}\frac{\Im\chi_\mathrm{d}(\bq,\omega+\ui\epsilon)}{1-\ue^{-\hbar\omega/k_\mathrm{B}T}}
\;.
\end{align}

When we set $U_{\B,\F}^\ext=0$, the coupled moment equations become an eigenequation.
Its eigenvalue is a complex, in general, 
the real part of which corresponds to the collective mode frequency $\omega$, and the imaginary part of which corresponds to the minus sign of the damping rate $\Gamma$ of the collective mode.

Collective excitations are characterized by frequency $\omega$ of the collective mode and collisional relaxation times $\tau_{\B,\F}$ to reach local equilibrium.
The possible regimes include:
(I) the hydrodynamic regime in which the collision rates are so high that both bosons and fermions are in local equilibrium ($\omega\tau_\B\ll1,\,\omega\tau_\F\ll1$).
In ordinary systems, the sound mode in the hydrodynamic regime is called first sound;
(II) the collisionless regime in which $\cI_{\B,\F}$ can be neglected ($\omega\tau_\B\gg1,\,\omega\tau_\F\gg1$).
In a Fermi gas, the sound mode in the collisionless regime is called zero sound;
(III) the intermediate regime in which collisions between bosons are sufficiently rapid to establish local equilibrium within a Bose gas, but collisions between bosons and fermions can be neglected, so that fermions are in the collisionless regime ($\omega\tau_\B\ll1,\,\omega\tau_\F\gg1$);
(IV) the regime opposite to III, where bosons are in the collisionless regime but fermions are in the hydrodynamic regime ($\omega\tau_\B\gg1,\,\omega\tau_\F\ll1$).
In these extreme limits, one can derive analytical expressions for the density response function and hence determine the collective mode frequency.
Of course, there is a region that cannot be categorized into one of the four regimes.
In the present paper, we apply the moment method to the Boltzmann equations, which can capture the crossover among the different regimes.
The moment method was successfully used to discuss the crossover between hydrodynamic regime and collisionless regime in a two component Fermi gas \cite{watabe_2010_JLTP_158_zero,narushima_2018_JPBAMOP_51_density}.

\section{Results}
\label{sec:results}
\begin{figure}
\begin{minipage}[c]{.40\textwidth}
\raggedright
\includegraphics[scale=.48]{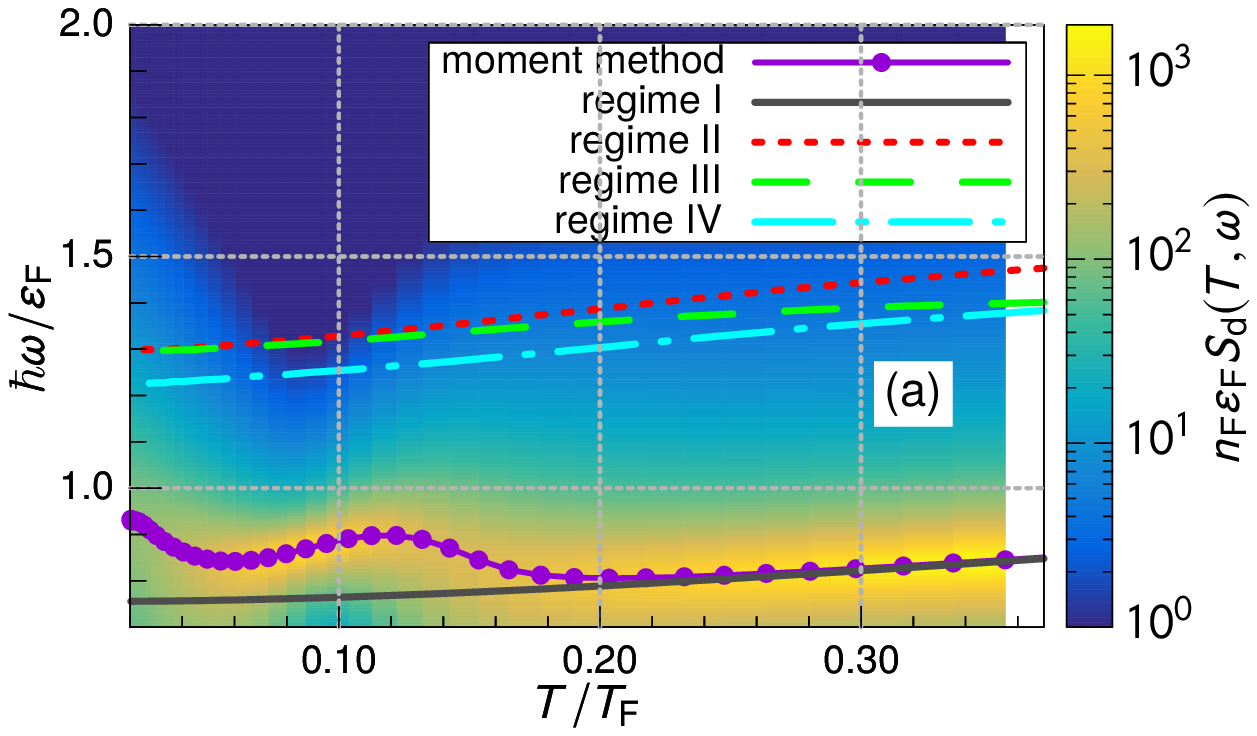}
\phantomsubcaption
\label{subfig:frequencies_and_structure_factor:1}
\\
\includegraphics[scale=.48]{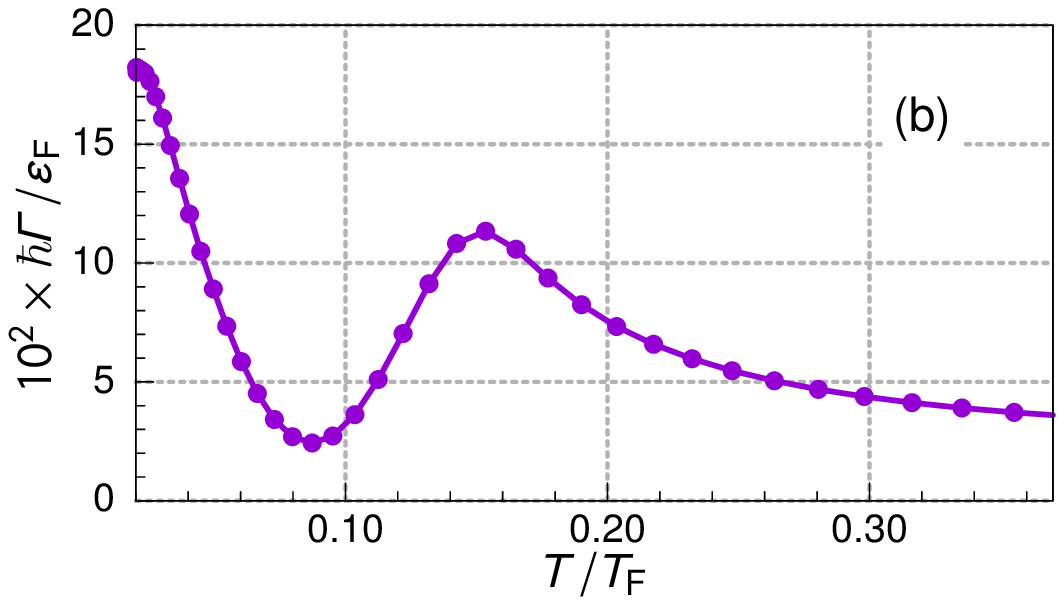}
\phantomsubcaption
\label{subfig:damping_rate:1}
\end{minipage}
\qquad
\begin{minipage}[c]{.40\textwidth}
\raggedright
\includegraphics[scale=.48]{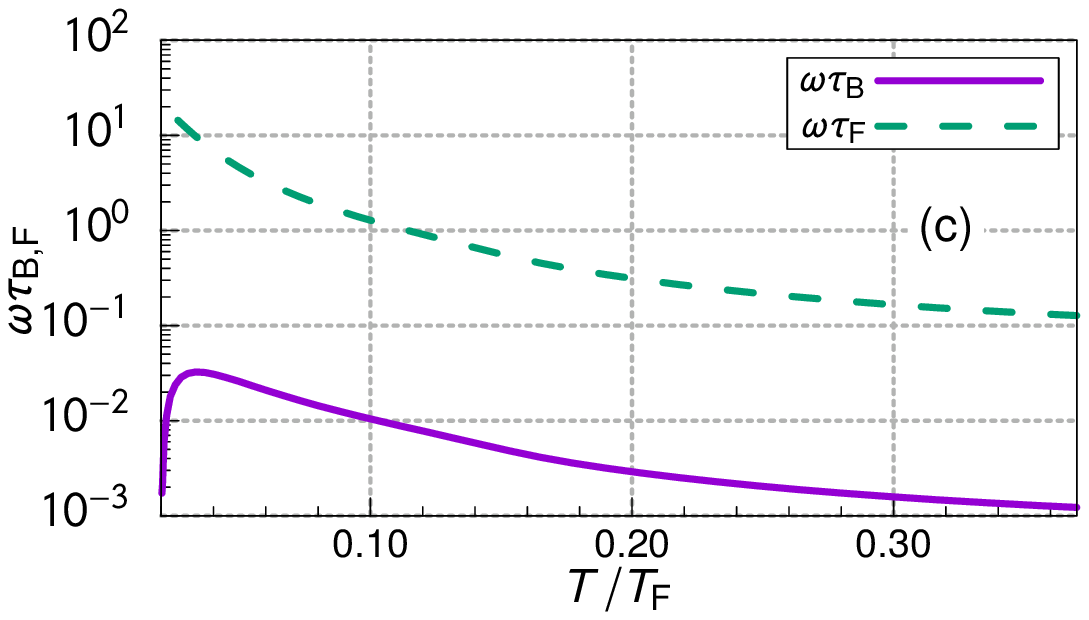}
\phantomsubcaption
\label{subfig:frequency_times_average_relaxation_time:1}
\\
\includegraphics[scale=.48]{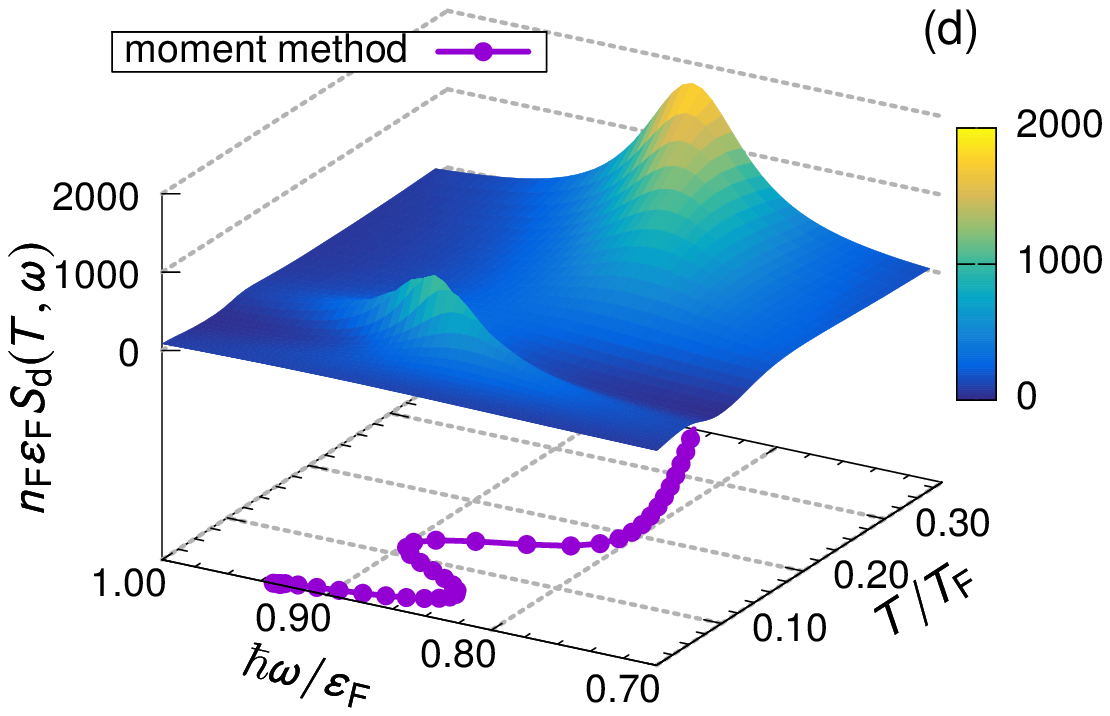}
\phantomsubcaption
\label{subfig:frequencies_and_structure_factor_3d:1}
\end{minipage}
\caption{
Results obtained by solving the coupled moment equations for $g_\BB=g_\BF=24\vepsF/n_\F$, $m_\B=m_\F$, and $n_\B/n_\F=0.01$.
Here, $\vepsF$ and $\kF$ denote Fermi energy and wave number, respectively, and $n_\F$ indicates equilibrium density of fermions.
\textbf{a}
False color plot of the dynamic structure factor with a fixed wave number $q=0.5\kF$ as a function of the frequency and the temperature.
We also plot the temperature dependence of the collective mode frequency $\omega$ 
obtained from the coupled moment equations, as well as the approximate mode frequency in the four limiting cases (I)--(IV).
\textbf{b} Damping rate $\Gamma$ of the collective mode.
\textbf{c} Temperature dependence of $\omega\tau_\B$ and $\omega\tau_\F$.
\textbf{d} Surface plot of the dynamic structure factor as a function of the frequency and the temperature (Color figure online).
}
\label{fig:result:1}
\end{figure}
We consider the normal Bose--Fermi mixture gas in the quantum degenerate regime.
To be in this regime, the Bose--Einstein condensation temperature $T_\BEC$ and the Fermi temperature $T_\F$ must satisfy $T_\BEC\leq T \ll T_\F$.
For ideal Bose and Fermi gases, $T_\BEC\propto n_\B^{2/3}/m_\B$ and $T_\F\propto n_\F^{2/3}/m_\F$, where $n_\B$ and $n_\F$ are equilibrium densities of bosons and fermions.
Thus, one has the relation $T_\BEC/T_\F\propto(m_\F/m_\B)(n_\B/n_\F)^{2/3}\ll1$.
In the present paper, we consider the case $n_\B/n_\F\ll1$.

Figure~\ref{fig:result:1} summarizes the main results obtained by solving the coupled moment equations for a given temperature $T$, imbalance parameters of population $n_\B/n_\F$ and mass $m_\B/m_\F$, interaction strengths of boson--boson $g_\BB$ and boson--fermion $g_\BF$, and wave number $|\bq|$. 
In this study, we discuss the temperature dependence of the collective mode, focusing on the frequency, damping rate, and dynamic density structure factor. 

Figure~\ref{subfig:frequencies_and_structure_factor:1} shows that
the dynamic structure factor as a function of the frequency and temperature.
We also plot the temperature dependence of the collective mode frequency obtained by solving the eigenvalue equation by setting $U^\ext_{\B,\F}=0$ in the coupled moment equations.
For comparison, we also plot the approximate mode frequency obtained by supposing that one is in the four limiting cases (I), (II), (III), and (IV).
It is seen that the peak strength of the dynamic structure factor increases not only at high temperature but also at low temperature
(see also Fig.~\ref{subfig:frequencies_and_structure_factor_3d:1}).
In the high temperature region, the collective mode frequency obtained from the coupled moment equations coincides in the hydrodynamic regime (I). 
With decreasing temperature, the collective mode frequency deviates from that of the first sound mode.
At low temperature much lower than $T_\F$, however, this frequency also deviates from the frequency in the other three limiting cases (II)-(IV), where either or both of bosons and fermions are in the collisionless regime. 
The peak of the density dynamic structure factor in the low temperature regime indicates that the long-lived collective mode emerges at $T\sim0.09T_\F$, which is different from the modes that one can expect from the four limiting cases (I), (II), (III), and (IV). 

Indeed, the damping rate $\Gamma$ of the collective mode takes a minimum at $T\sim0.09T_\F$, where the peak of the dynamic structure factor increases (Fig.~\ref{subfig:damping_rate:1}). 
Here, the damping rate $\Gamma$ is obtained from the imaginary part of an eigenvalue of the coupled moment equations when $U^\ext_{\B,\F}=0$.
From these aspects, it is remarkable that there is a definite collective mode with the very small damping rate in the low temperature, which is different from four limiting cases we have assumed above. 
We also find that the damping rate increases with decreasing temperature in the low temperature regimes at $T\lesssim0.09T_\F$. 
The normal Bose--Fermi gas cannot hold the definite collective mode in the very low temperature regime close to the critical temperature of the Bose--Einstein condensation.  

We discuss what the region is for the collective mode, where the definite collective mode emerges at $T\sim0.09T_\F$, and where the system is close to the critical temperature of the Bose--Einstein condensation. 
In Fig.~\ref{subfig:frequency_times_average_relaxation_time:1}, we plot $\omega\tau_{\B,\F}$, which indicates that if $\omega\tau_{\B(\F)}\ll1$, the collective mode is in the hydrodynamic regime for bosonic (fermionic) species, and if $\omega\tau_{\B(\F)}\gg1$, the collective mode is in the collisionless regime for bosonic (fermionic) species. 
For the fermions, the collective mode at the high temperature is in the hydrodynamic regime, 
and that in the very low temperature regime is relatively in the collisionless regime, which is consistent with the normal fermionic system~\cite{watabe_2010_JLTP_158_zero,narushima_2018_JPBAMOP_51_density}. 
As in the case of the two-component Fermi gas and the Fermi liquid, the collision between boson and fermion is also suppressed in the very low temperature regime, because of the Pauli-blocking of the fermions.  
For the bosons, the collective mode is in the hydrodynamic regime over the whole temperature range. 
Although $\omega\tau_\B$ increases with the system being in the low temperature regime, we find that $\omega\tau_\B$ sharply drops at the temperature very close to the critical temperature of the Bose--Einstein condensates. 
It indicates that the collective mode near the critical temperature is deeply in the hydrodynamic regime again. 
This is because at the critical temperature, the Bose distribution function shows the infrared divergence, and the mean-collision time drastically shortens near the critical temperature. 

For the long-lived collective mode at the low temperature $T\sim0.09T_\F$, although the collective mode is in the hydrodynamic regime for bosons where $\omega\tau_\B\ll1$, it is in the crossover regime for fermions where $\omega\tau_\F\sim1$. 
According to our knowledge on the normal Fermi liquid theory, the damping rate is very large in the crossover regime. 
In this sense, this long-lived collective mode around $T\sim0.09T_\F$ is a very interesting excitation in the normal Bose--Fermi mixture gas. 
At the temperature very close to the critical temperature, the system is relatively in the limiting case (III), where the fermions are in the collisionless regime, and the bosons are in the hydrodynamic regime. 
However, we find that in this regime, the system cannot hold any definite long-lived collective mode from our numerical calculation.

\section{Conclusions}
\label{sec:conclusions}
The dynamic properties of the Bose--Fermi mixtures in normal state are studied by using the moment method.
In order to consider the quantum degenerate regime at very low temperature and yet in the normal state, we assumed that a number ratio of bosons to fermions $n_\B/n_\F$ is very low.
We find that, in the low temperature region much lower than the Fermi temperature, the damping rate takes a minimum as a function of the temperature, and the long-lived collective sound mode emerges, which is different from the conventional first and zero sounds. 
From our numerical calculation, although this long-lived excitation is regarded as the collective mode in the hydrodynamic regime for bosons, 
it is regarded as the collective mode in the crossover regime between the hydrodynamic and collisionless regimes for fermions.

\section*{Acknowledgements}
We are grateful to Y.~Iijima for discussion in the early stage of this work.
S.W. is supported by JSPS KAKENHI Grant No. (JP16K17774, JP18K03499),
and T.N. is supported by JSPS KAKENHI Grant No. (JP16K05504).

\bibliographystyle{spphys}       
\bibliography{bose_fermi_mixtures_preprint}   
\end{document}